\documentclass[12pt,fleqn]{article}
\usepackage[dvipdfmx]{graphicx}
\usepackage{wrapfig}
\usepackage{amsmath,amssymb}
\usepackage{color}
\usepackage{xcolor}
\usepackage[breaklinks=true]{hyperref}
\usepackage{ulem}
\usepackage{cite}
\begin{document}
\baselineskip 0.6cm
\renewcommand{\thefootnote}{\#\arabic{footnote}} 
\setcounter{footnote}{0}
%
\begin{titlepage}
\begin{center}

\begin{flushright}
\end{flushright}


{\Large \bf 
On radiative corrections to 
lepton number violating processes
}\\

\vskip 1.2cm

{\large 
Takehiko Asaka$^1$,
Hiroyuki Ishida$^2$,
and
Kazuki Tanaka$^3$
}

\vskip 0.4cm

$^1${\it
  Department of Physics, Niigata University, Niigata 950-2181, Japan
}

$^2${\it
Center for Liberal Arts and Sciences, Toyama Prefectural University, Toyama 939-0398, Japan
}

$^3${\it
  Graduate School of Science and Technology, Niigata University Niigata, 950-2181, Japan
}

\vskip 0.2cm

(December 10, 2024)

\vskip 2cm

\begin{abstract}
	We consider the minimal model of the seesaw mechanism by introducing two right-handed neutrinos,  whose masses are comparable to
	the electroweak scale. 
    This framework is attractive, since it is testable at terrestrial experiments. 
    A critical consequence of this mechanism is the violation of lepton number conservation due to the Majorana masses of both active neutrinos and heavy neutral leptons.
	In particular, we investigate the impact of the radiative corrections to Majorana masses of
	left-handed neutrinos on the lepton number violating processes, such as the neutrinoless double beta
	decay: $(Z, A) \to (Z+2,A) + 2 e^-$ and the inverse neutrinoless double beta decay: $e^- e^- \to W^- W^-$.
	It is shown that the cross section of the inverse neutrinoless double beta decay
	can increase by ${\cal O}(10)$~\% when the masses of heavy neutral leptons 
	are ${\cal O}(1)$~TeV, which has significant implications on future experiments.
\end{abstract}
\end{center}
\end{titlepage}

\section{Introduction}
The neutrino masses are revealed by various oscillation experiments and its origin remains one of the most significant open problems in particle physics. 
Among the numerous theoretical scenarios proposed so far, the seesaw mechanism, 
which introduces right-handed neutrinos~\cite{Minkowski:1977sc,Yanagida:1979as,Yanagida:1980xy,Ramond:1979,GellMann:1980vs,Glashow:1979},
is widely regarded as a particularly compelling framework.
This is because it naturally explains the observed smallness of neutrino masses.
It is commonly discussed that the energy scale of the seesaw mechanism, 
specifically the scale of the Majorana masses of right-handed neutrinos, is extremely high, 
often approaching the scale of the gauge unification. 
However, the seesaw scale can also be much lower than such a conventional scale, 
potentially comparable to or even below the electroweak scale $\sim \mathcal{O}(10^2)$~GeV. 
A low-scale seesaw mechanism is especially intriguing, 
as it opens up the possibility of experimental verification through terrestrial experiments~\cite{Gorbunov:2007ak,Atre:2009rg,Deppisch:2015qwa}.

In the seesaw mechanism, 
the neutrino mass eigenstates consist of active neutrinos and heavy neutral leptons (HNLs), 
both of which are Majorana fermions. 
The presence of Majorana masses inherently violates the lepton number conservation, 
leading to the possibility of processes that are forbidden within the Standard Model (SM). 
These processes provide a unique window to probe the nature of neutrinos and the underlying mechanism of their mass generation.
The well-known example is the neutrinoless double beta ($0\nu \beta \beta$) decay: $(Z, A) \to (Z+2,A) + 2 e^-$~\cite{Furry:1939qr}, which breaks the lepton number by two units.
The contribution to the decay from active neutrinos and HNLs is parameterized by
the so-called effective mass $m_{\rm eff}$.  
The upper bound at present is $|m_{\rm eff}| < $28--122~meV~\cite{KamLAND-Zen:2024eml} where
the range denotes the uncertainty in estimating the nuclear matrix element of the process.

Another intriguing lepton number violation (LNV) process is
the inverse neutrinoless double beta (i$0 \nu \beta \beta$) decay:
$e^- e^- \to W^- W^-$~\cite{Rizzo:1982kn}.
This is a promising target for the future $e^- e^-$ collision experiments at high-energy facilities,
such as the International Linear Collider (ILC)~\cite{Baer:2013cma}  and the Compact Linear Collider (CLIC)~\cite{Accomando:2004sz,Brunner:2022usy}.
Various analyses of this process have been carried in the literature so 
far~\cite{Rizzo:1982kn,London:1987nz,Dicus:1991fk,Gluza:1995ix,%
Belanger:1995nh,Gluza:1995js,Greub:1996ct,Rodejohann:2010jh,Banerjee:2015gca,Asaka:2015oia}.
This process offers advantages over the $0 \nu \beta \beta$ decay, as its signal is clean 
and its theoretical predictions are free from uncertainties in the nuclear matrix elements. 
Consequently, it provides a complementary and robust test for the LNV.

The contribution to these LNV processes from HNLs can be sizable when their mixing
in the charged current interaction becomes large considerably.
In such cases, the tuning among parameters of neutrinos is required
in order to reproduce the observed masses of active neutrinos.
Thus, it is essential to pay close attention to the radiative corrections to the seesaw mechanism,
to be specific, the corrections to the Majorana masses of left-handed 
neutrinos~\cite{Pilaftsis:1991ug,Grimus:2002nk,AristizabalSierra:2011mn},
which are absent at the tree-level.
Such corrections to the $0 \nu \beta \beta$ decay have been discussed in Ref.~\cite{Lopez-Pavon:2015cga}.
We will, thus, investigate the impact of the radiative corrections to the
i$0 \nu \beta \beta$ decay in the present analysis.

The structure of this paper is as follows: 
In Sec.~2, we briefly explain the seesaw mechanism and review the radiative corrections to it. In particular, we present the parametrization of the neutrino Yukawa matrix, including the effects of these corrections~\cite{Lopez-Pavon:2015cga}.
In Sec.~3, we study the impacts of the radiative corrections on the LNV processes.
In particular, we perform a numerical estimation of the cross section for $e^- e^- \to W^- W^-$ and discuss the quantitative significance of these corrections. 
Finally, in Sec.~4, we summarize our results and conclusions. 

\section{Radiative corrections to the seesaw mechanism}\label{sec:Model}
We begin by introducing the model used in the present analysis.
We consider the SM extended by two right-handed neutrinos $\nu_{RI} $ ($I=1,2$),
which represents the minimal framework capable of explaining the two mass-squared differences observed in neutrino oscillation experiments.
The Lagrangian of this model is given by:
\begin{align}
	{\cal L}
	= {\cal L}_{\rm SM} 
	+
  \overline \nu_{R I} i \partial_\mu \gamma^\mu   \nu_{R I}
  - \left( F_{\alpha I} \overline L_\alpha H \nu_{R I}
  + \frac{M_I}{2} \overline{\nu_{R I}^c} \nu_{R I} 
  + h.c. \right) \,,
\end{align}
where $H$ is Higgs doublet and
$L_\alpha = (\nu_{L\alpha}, \ell_{L\alpha})^T$~($\alpha = e, \mu, \tau)$ 
are left-handed lepton doublets.
The mass matrix at tree-level is 
\begin{align}
	\hat M^{(0)}
	=
	\left(
		\begin{array}{c c}
			0 & M_D \\ M_D^T & M_M 
		\end{array}
	\right) \,,
\end{align}
where Dirac mass matrix is $M_D = F \langle H \rangle$ and 
Majorana mass matrix is $M_M  = \mbox{diag}( M_1, M_2)$.
Assuming a hierarchy between
Dirac masses and Majorana masses, $|M_D|_{\alpha I} \ll M_M$,
the seesaw mechanism is implemented.
In this framework, the lighter mass eigenstates correspond to active neutrinos, $\nu_i$ ($i=1,2,3$),
with masses $m_i$, while the heavier states correspond to heavy neutral leptons (HNLs), $N_I$, with masses $M_I$. 
It is important to note that the lightest active neutrino is massless, as the model introduces only two right-handed neutrinos.
For simplicity, we focus on the normal hierarchy of active neutrino masses,
and then $m_3 > m_2 > m_1=0$.

To express the mixing elements of HNLs, $\Theta = M_D/M_M$,
we utilize the Casas-Ibarra parametrization of the neutrino Yukawa couplings~\cite{Casas:2001sr}.
At the tree-level, the mass matrix of active neutrinos is obtained as
\begin{align}
	\label{Mnu0}
	M_\nu^{(0)}
	= - M_D M_M^{-1} M_D^T \,,
\end{align}
which gives the tree-level expression of the mixing elements 
\begin{align}
	\Theta_{\alpha I}^{(0)}
	=
	i \, \left[ U D_\nu^{1/2} \Omega M_M^{1/2} M_M^{-1} \right]_{\alpha I} \,,
\end{align}
where $U$ denotes the PMNS matrix and $D_\nu = \mbox{diag}(m_1=0, m_2, m_3)$.
The $3 \times 2$ orthogonal matrix $\Omega$ can be taken in the form
\begin{align}
	\label{OM}
  \Omega =
  \left(
    \begin{array}{c c}
      0 & 0 \\
      \cos \omega & - \sin \omega \\
      \sin \omega & \cos \omega
    \end{array}
  \right) \,,
\end{align}
where $\omega$ is a complex parameter.
We shall introduce a parameter $X_\omega$ as 
\begin{align}
	X_\omega = \exp \left( \mbox{Im} \omega \right) \,,
\end{align}
which will be crucial for quantifying the magnitudes of the neutrino Yukawa couplings 
namely, the magnitude of the mixing elements.

We now turn our attention to the radiative corrections to the seesaw mechanism.
The relevant corrections at one-loop level are those to Majorana masses 
of left-handed neutrinos $\nu_{L \alpha}$~\cite{Pilaftsis:1991ug,Grimus:2002nk,AristizabalSierra:2011mn}:
\begin{align}
	\delta M_{LL}
	=
	M_D M_M^{-1} \delta_{LL} M_D^T \,,
\end{align}
where $\delta_{LL}
	=
	\mbox{diag}
	\left( f_{\delta_{LL}} (M_1) , f_{\delta_{LL}} (M_2) \right)$ 
and the function $f_{\delta_{LL}}$ is given by	
\begin{align}
	f_{\delta_{LL}} (M)
	=
	\frac{M^2}{(4 \pi \langle H \rangle)^2}
	\left[ 
		\frac{3 \ln \left( \frac{M^2}{m_Z^2} \right)}
		{ \frac{M^2}{m_Z^2} - 1 }
		+
		\frac{\ln \left( \frac{M^2}{m_H^2} \right)}
		{ \frac{M^2}{m_H^2} - 1 }
	\right] \,.
\end{align}
The mass matrix of active neutrinos and HNLs at one-loop level
is then given by
\begin{align}
	\hat M^{(1)} =
	\left(
		\begin{array}{c c}
			\delta M_{LL} & M_D \\ M_D^T & M_M
		\end{array}
	\right) \,.
\end{align}
Here we have neglected the corrections to the Dirac masses $M_D$ and the Majorana masses $M_M$,
since they are subdominant due to the loop suppression.
In this case the mass matrix of active neutrinos at one-loop level is
\begin{align}
 M_\nu^{(1)} =  M_\nu^{(0)} + \delta M_{LL} 
	= - M_D \tilde M_M^{-1} M_D^T \,.
\end{align}
where the diagonal matrix $\tilde M_M$ is given by
\begin{align}
	&\tilde M_M = M_M ( 1 + \delta_{M_M} ) \,,
	\\
	&\delta_{M_M} = 	\delta_{{LL}} ( 1- \delta_{{LL}})^{-1} \,.\label{delMM}
\end{align}
Here, since $\delta_{M_M}$ is also diagonal matrix, $\tilde{M}_M$ is still diagonal matrix after including one-loop corrections.
It is found from Eq.~(\ref{Mnu0}) that
$M_\nu^{(1)}$ is obtained from $M_\nu^{(0)}$
replacing $M_M$ by $\tilde M_M$.
Consequently, the Casas-Ibarra parametrization of the HNL's mixing elements
is~\cite{Lopez-Pavon:2015cga}
\begin{align}
	\Theta^{(1)}_{\alpha I}
	=
	i \left[ U D_\nu^{1/2} \Omega \tilde M_M^{1/2}  M_M^{-1}
	\right]_{\alpha I} \,.
\end{align}

Therefore, the radiative corrections to the seesaw mechanism
is represented by $\delta_{M_M}$.
\begin{figure}[t]
  \centerline{
  \includegraphics[width=10cm]{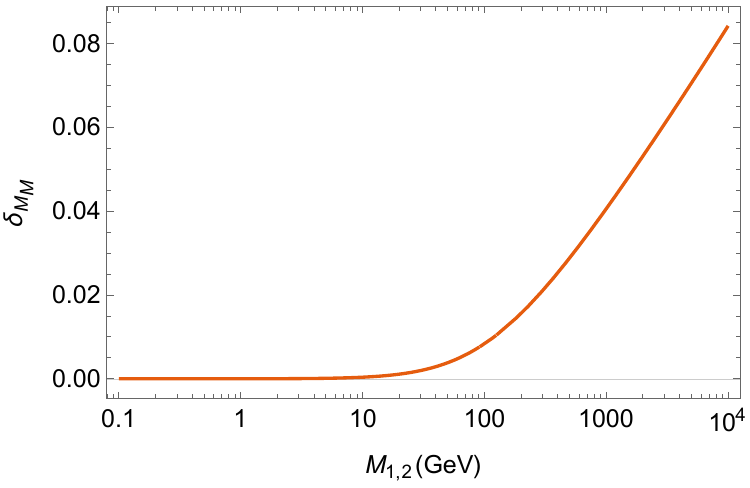}
  }%
  \caption{
    The behavior of the function $\delta_{M_M}$ 
    as a function of $M_{1,2}$.
    The values of $m_Z$ and $m_H$ are taken from~Ref.~\cite{ParticleDataGroup:2024cfk}.
  }
  \label{F_DMM}
\end{figure}
As shown in Fig.~\ref{F_DMM},
the effect of $\delta_{M_M}$ is negligible for $M_{1,2} < {\cal O}(10)$~GeV
while it becomes a sizable value for $M_{1,2} \gg {\cal O}(10)$~GeV.
In the next section, we will estimate numerically the significance of these corrections to 
the LNV processes.

\section{Corrections to lepton number violating processes}
Within the framework of the seesaw mechanism, both active neutrinos and HNLs are Majorana particles. The presence of Majorana masses inherently violates the lepton number conservation, giving rise to distinctive processes that are forbidden in the SM.
We discuss here two well-known examples, the $0\nu \beta \beta$ decay
and the i$0 \nu \beta \beta$ decay.
Especially, we investigate the quantitative significance 
of the radiative corrections to the i$0 \nu \beta \beta$ decay.

First, we consider the $0\nu \beta \beta$ decay.  This process,
which violates the lepton number by two units, can be induced by massive Majorana neutrinos.
In the scenario under consideration, both active neutrinos and HNLs contribute to the process.
The half-life of the $0\nu \beta \beta$ decay for a given nucleus is expressed as
\begin{align}
	(T_{1/2}^{0\nu \beta \beta})^{-1} = G^{ 0\nu \beta \beta}
	|M^{0\nu \beta \beta}|^2 | m_{\rm eff} |^2\,, 
\end{align}
where $G^{0\nu \beta \beta}$ is the phase space factor and
$M^{0\nu \beta \beta}$ is the nuclear matrix element of the process.
The so-called effective mass $m_{\rm eff}$ represents 
the effect of massive Majorana neutrinos, which is given by
\begin{align}
	m_{\rm eff} 
	= m_{\rm eff}^\nu + m_{\rm eff}^N 
	=
	\sum_{i} m_i U_{ei}^2 + \sum_I M_I \Theta_{eI}^2 \, f_\beta (M_I) \,,
\end{align}
where the first term $m_{\rm eff}^\nu$ denotes the contribution from active neutrinos
and the second $m_{\rm eff}^N$ from HNLs.  
The nuclear matrix elements for HNLs are suppressed compared
to those of active neutrinos when their masses are heavier than $\sim 100$~MeV.
This effect is encapsulated in the function $f_\beta$.
The precise form of $f_\beta$ requires a detailed calculation 
of the nuclear matrix element, but here we use the following approximate form
(see Ref.~\cite{Faessler:2014kka})
\begin{align}
	f_\beta (M_I) = \frac{\langle p^2 \rangle}{\langle p^2 \rangle + M_I^2} \,,
\end{align}
where
$\sqrt{\langle p^2 \rangle} = {\cal O}(100)$~MeV is the typical Fermi momentum.
As a representative value, we take $\sqrt{\langle p^2 \rangle} = 200$~MeV from now on.

To begin with, we discuss the case without the radiative corrections to neutrino 
masses, namely $\delta M_{LL}=0$. 
The effective mass in this case is estimated as
\begin{align}
	m_{\rm eff}^{(0)}
	=
	- \big[
		U D_\nu^{1/2} \Omega \delta_{f_\beta} \Omega^T D_\nu^{1/2} U^T
	\big]_{ee} \,,
\end{align}
where the $2 \times 2$ matrix $\delta_{f_\beta}$ is defined by
\begin{align}
	\delta_{f_\beta}
	=
	\mbox{diag} \big( f_\beta (M_1) -1 , f_\beta (M_2) - 1\big)
	=
	\mbox{diag} \left( - \frac{M_1^2}{\langle p^2 \rangle + M_1^2} ,
	- \frac{M_2^2}{\langle p^2 \rangle + M_2^2} \right) \,.
\end{align}
It has been shown that,
when $M_{1,2} \ll \sqrt{\langle p^2 \rangle}$,
$m_{\rm eff}^{(0)}$ vanishes since $\delta_{f_\beta} \to 0$.
This is a direct consequence of $\delta M_{LL}=0$ in the tree-level relation of the seesaw mechanism.
On the other hand, when $M_{1,2} \gg \sqrt{\langle p^2 \rangle}$,
HNLs decouple from the $0 \nu \beta \beta$ decay and 
$m_{\rm eff}^{(0)} \simeq m_{\rm eff}^\nu$.
This is because
$\delta_{f_\beta} \simeq - 1$ for $M_{1,2}^2/\langle p^2 \rangle \gg 1$
and then $m_{\rm eff}^{(0)} \simeq [U D_\nu^{1/2} \Omega \Omega^T D_\nu^{1/2} U^T ]_{ee}
= [U D_\nu U^T]_{ee} = m_{\rm eff}^\nu$.

Next, we will examine the impact of radiative corrections taking $\delta M_{LL} \neq 0$.
Although it has already been discussed in the literature (for example, see Ref.~\cite{Lopez-Pavon:2015cga}), we revisit this 
issue in order to derive the explicit consequences from $m_{\rm eff} =0$, namely, the cancellation in $m_{\rm eff}$ between
active neutrinos and HNLs, by taking into account the radiative corrections.
The non-zero corrections $\delta M_{LL}$ modifies the effective mass as
\begin{align}
	\label{meff1}
	m_{\rm eff}^{(1)}
	=
	m_{\rm eff}^{(0)} + \delta m_{\rm eff}^{(1)} \,,
\end{align}
where the correction is given by
\begin{align}
\label{dmeff1}
	\delta m_{\rm eff}^{(1)}
	=
	- \big[
		U D_\nu^{1/2} \Omega \delta_{M_{M}} ( 1 + \delta_{f_\beta} )
		\Omega^T D_\nu^{1/2} U^T
	\big]_{ee} \,.
\end{align}
Note that $\delta_{M_{M}}$ is defined in Eq.~(\ref{delMM}).

\begin{figure}[t]
  \centerline{
  \includegraphics[width=10cm]{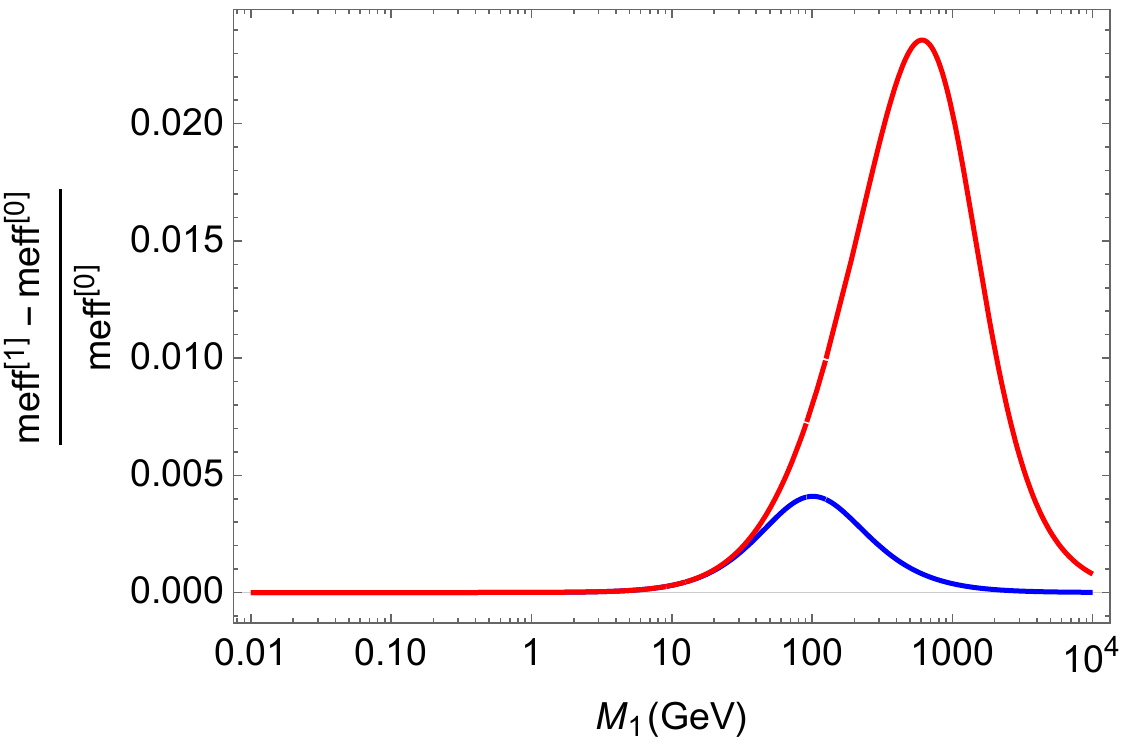}
  }%
  \caption{
  	Impacts of the radiative corrections to the effective mass of
	the $0 \nu \beta \beta$ decay for $X_\omega = 10^4$ (red line)
	and $10^3$ (blue line).
	We take $M_2 = 10 M_1$ and Re$\omega = \pi/4$.
  }
  \label{F_meffcomp}
\end{figure}
The impact of the radiative corrections on $m_{\rm eff}$ is shown in Fig.~\ref{F_meffcomp}.
For the chosen parameters, the corrections are less than a few percent.
When $M_{1,2}$ are sufficiently lighter than $m_{H,W}$,
$\delta_{M_M}$ becomes negligible, and hence the correction $\delta m_{\rm eff}^{(1)}$ vanishes.
On the other hand, when $M_{1,2}^2/\langle p^2 \rangle \gg 1$,
$\delta_{f_{\beta}}$ becomes close to $-1$ and then $\delta m_{\rm eff}^{(1)}$ vanishes.

It should be noted that, when we take $X_\omega \gg 1$,
the effective mass becomes large, since $|\Theta_{e I}|$ becomes proportional to
$X_\omega$ in such cases.
See, for example, the discussion in Ref.~\cite{Asaka:2011pb}.

When $M_1 = 5 \times 10^3$ GeV, $M_2 = 10^2 M_1$, and Re$\omega =\pi/4$,
$|m_{\rm eff}| < 122$~(28)~meV leads to the upper bound on $X_\omega$:
$X_\omega < 2.84~(1.42) \times 10^5$ at tree level and
$X_\omega < 2.75~(1.38) \times 10^5$ at one-loop level, respectively.%
\footnote{
In the present analysis we adopt the central values of mixing angles, mass-squared differences, and CP-violating phase of active neutrinos in Ref.~\cite{Esteban:2024eli}.
}
This shows that the loop correction modifies the upper bound on $X_\omega$ about 3~\%.
In addition, the impact of the radiative corrections on $m_{\rm eff}$ of the $0\nu\beta\beta$ decay appears to be less significant 
and is likely obscured by the uncertainties in the nuclear matrix element.

\begin{figure}[t]
  \centerline{
  \includegraphics[width=10cm]{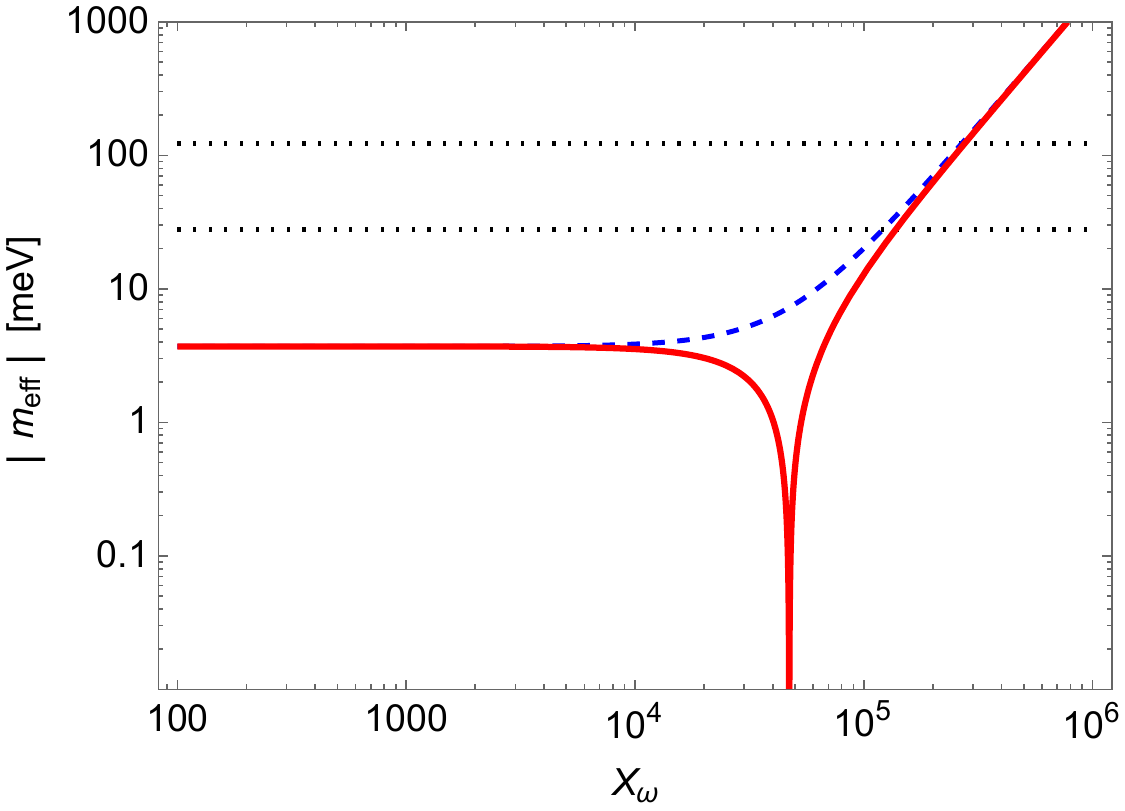}
  }%
  \caption{
  	Effective mass $m_{\rm eff}$ including the radiative corrections
	for Re$\omega =$ Re$\omega_+$ (red line) and Re$\omega = \pi/4$ (blue line).
	We take $M_1=5$~TeV and $M_2= 10^2 M_1$.
	The horizontal dotted lines show the upper bounds on
	$|m_{\rm eff}| = 28$~meV and 122~meV~\cite{KamLAND-Zen:2024eml}.
	  }
  \label{F_Cmeff}
\end{figure}

The authors of the present analysis have pointed out the
possibility that $m_{\rm eff} =0$ due to the cancellation between
the contributions of active neutrinos and HNLs~\cite{Asaka:2020wfo,Asaka:2020lsx}.
We demonstrate how the condition of $m_{\rm eff} =0$ is modified
by the inclusion of the radiative corrections.
It is found from Eqs.~(\ref{meff1}) and (\ref{dmeff1}) that
$m_{\rm eff} =0$ is realized if the parameter $\omega$ in Eq.~(\ref{OM}) is
\begin{align}
	\tan \omega = \tan \omega_\pm \equiv \frac{A \pm i \tilde \delta}{1 \mp i A \tilde \delta} \,
\end{align}
where the parameter $A$ is the same as the one defined in Refs.~\cite{Asaka:2020wfo,Asaka:2020lsx}:
\begin{align}
	A = \frac{U_{e3} m_3^{1/2}}{U_{e2} m_2^{1/2}} \,,
\end{align}
where the result is for the NH case.
On the other hand, $\delta_f$ in Eq.~(21) of \cite{Asaka:2020lsx} is replaced by 
$\tilde \delta$ which is given by
\begin{align}
	\tilde \delta = \sqrt{ \frac{[\delta_{f_\beta} + \delta_{M_M}(1+\delta_{f_\beta})]_{11}}
	{[\delta_{f_\beta} + \delta_{M_M}(1+\delta_{f_\beta})]_{22}}} \,.
\end{align}
This shows that the cancellation in $m_{\rm eff}$ is still possible even with
the radiative corrections to neutrino masses, which is shown in Fig.~\ref{F_Cmeff}.

~

Finally, we proceed to discuss another process of lepton number violation
\begin{align}
	e^- e^- \to W^- W^- \,,
\end{align}
which is called as the ``inverse neutrinoless double beta (i$0\nu \beta \beta$) decay".
This process violates the lepton number by two units, and can be induced by Majorana active neutrinos 
and HNLs in the seesaw mechanism~\cite{London:1987nz,Dicus:1991fk,Gluza:1995ix,Belanger:1995nh,Greub:1996ct,Rodejohann:2010jh,Banerjee:2015gca,Asaka:2015oia}.
When we consider the contribution only from active neutrinos,
the cross section is given by~\cite{Belanger:1995nh,Rodejohann:2010jh,Asaka:2015oia}
\begin{align}
	\sigma_\nu = \frac{3 G_F^2 |m_{\rm eff}^\nu|^2}{4 \pi}
	=  1.9 \times 10^{-19}
	\left( \frac{|m_{\rm eff}^\nu|}{122~\mbox{meV}}\right)^2 \,,
\end{align}
and $\sigma_\nu$ is too small to be probed by future colliders, 
which means that the i$0\nu \beta \beta$ decay offers a good test for HNLs 
if they give significant contributions to the decay.

Let us consider here the impact of the radiative corrections to neutrino masses
on the i$0\nu \beta \beta$ decay.   The cross section of the process in the considering model 
is expressed as (see, for example, Ref.~\cite{Asaka:2015oia})
\begin{align}
	\frac{d \sigma}{d \cos \theta}
	=
	\frac{G_F^2 \beta_W}{32 \pi}
	\Big[
		|A_t|^2 B_t + |A_u|^2 B_u + (A_t A_u^\ast + A_t^\ast A_u) B_{tu} \,,
	\Big]
\end{align}
where $\beta_W = \sqrt{1- 4 r_W}$ with $r_W = m_W^2/s$ (the center-of-mass energy $\sqrt{s}$).
The functions $A_{t,u}$ and $B_{t,u, tu}$ take the forms as
\begin{align}
	&A_t = \sum_{i = 1}^3 \frac{U_{ei}^2 \, m_i}{t - m_i^2} 
  + \sum_{I=1}^{2} \frac{\Theta_{eI}^2 \, M_I}{t - M_I^2}  \,,
  \\
  &A_u = \left. A_t \right|_{ t \to u} \,,
  \\
  &B_t = ( 1 - 4 \, r_W ) \, t^2 - 4 \, r_W ( 1- 2 \, r_W) \, s\, t  + \, 4 (  1 - r_W ) \, r_W^2 \, s^2 \,,
  \\
  &B_u = \left. B_t \right|_{t \to u} \,,
  \\
  &B_{tu} = ( 1 - 4 \, r_W) \, t \, u + 4 \, r_W^3 \, s^2 \,,
\end{align}
where the Mandelstam variables are defined as usual.
It is seen that the relative phases between the mixing elements,
$U_{ei}^2$ and $\Theta_{eI}^2$, are important to estimate $\sigma$~\cite{Gluza:1995ix,Gluza:1995js}.

When the masses of HNLs are at the TeV scale and their mixing elements are large,
we must take into account the constraints from the $0\nu \beta \beta$ decay
and the electroweak precision tests.
As for the former one, we impose
$|m_{\rm eff}| = |m_{\rm eff}^\nu + m_{\rm eff}^N| < 122$~meV~\cite{KamLAND-Zen:2024eml},
while the latter provides the upper bound on the mixing elements:
$| \Theta_{e1}|^2 + |\Theta_{e2}|^2 < 2.1 \times 10^{-3}$~\cite{Antusch:2014woa}.

\begin{figure}[t]
  \centerline{
  \includegraphics[width=10cm]{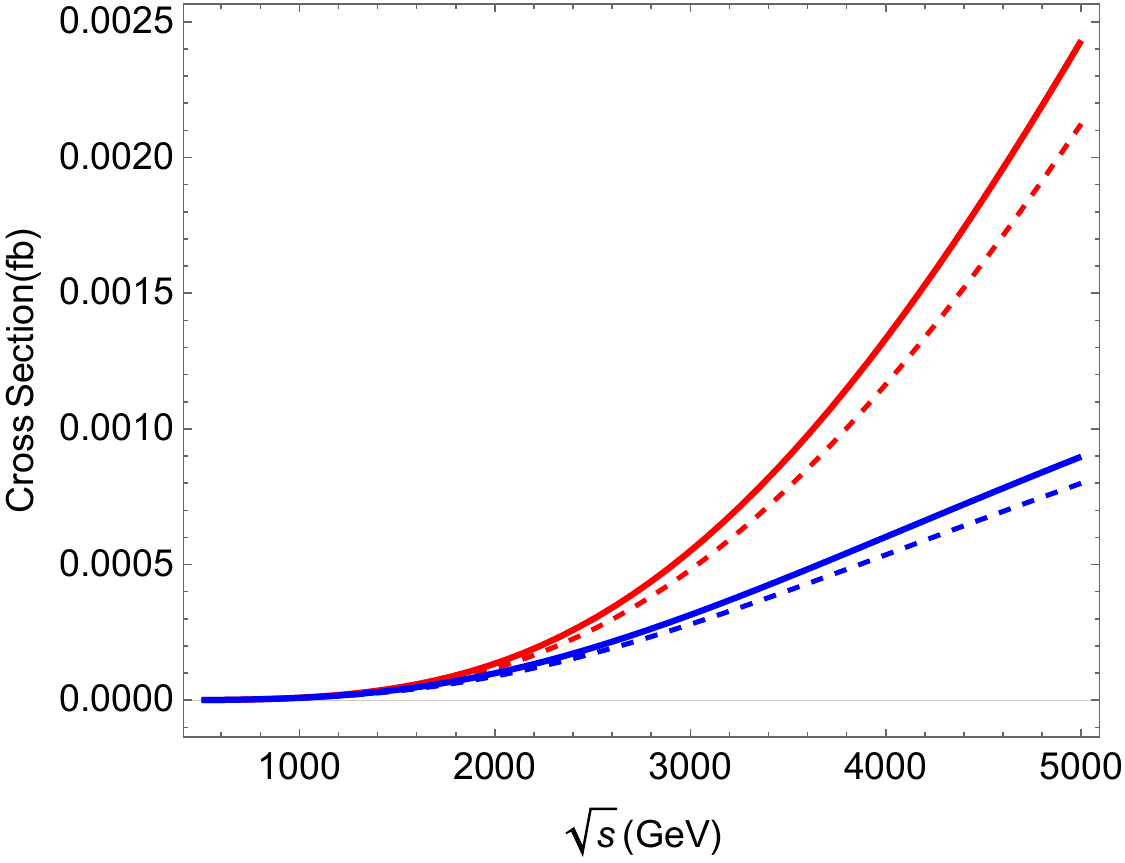}
  }%
  \caption{
	Cross sections of $e^- e^- \to W^- W^-$
	for $M_1 = 5$~TeV (red lines) and $3$~TeV (blue lines).
	We impose the constraints $|m_{\rm eff}| < 122$~meV and $|\Theta_{e}|^2 < 2.1 \times 10^3$.
	We take $M_2 = 10 M_1$. The solid and dashed lines correspond to the cases with and without
	the radiative correction to neutrino masses, respectively.
  }
  \label{F_CS_r10}
\end{figure}
Since the mixing elements of HNLs, $\Theta_{eI}$, are modified, 
we expect a correction to the cross section of $e^- e^- \to W^- W^-$.
In Fig.~\ref{F_CS_r10} we show the cross section of $e^- e^- \to W^- W^-$
with and without the radiative corrections to neutrino masses.
We estimate the maximal cross section by imposing the above constraints.
It is found that the cross section increases by about $15$~\% at $\sqrt{s} = 3$~TeV
due to the radiative correction to neutrino masses. 
This result has an impact on the search for HNL at the future experiments such as
the Compact Linear Collider (CLIC).

\section{Conclusions}
We have considered the minimal model of the seesaw mechanism
with two right-handed neutrinos, which masses are comparable to the electroweak scale $ \sim {\cal O
}(10^2)$ GeV.
This framework provides an opportunity to study heavy neutral leptons by experiments with accessible energy scales.
In this model, the lepton number is violated by the Majorana nature of active neutrinos
and heavy neutral leptons, and then the neutrinoless double beta decay as well as
the inverse neutrinoless double beta decay can be induced.
With suitable choice of the masses and mixing elements, 
heavy neutral leptons can significantly contribute to these processes.
We have examined the impacts of the radiative corrections to
the Majorana masses of left-handed neutrinos.
It has been shown that the cross section of $e^- e^- \to W^- W^-$
can increase by about 15~\% due to the corrections at $\sqrt{s} = 3$~TeV.
This enhancement holds significant implications for future experiments, such as the Compact Linear Collider (CLIC), 
in the search for the lepton number violation by heavy neutral leptons.

\section*{Acknowledgments}
This work was supported in part by JSPS KAKENHI Grant Number 24K07023 (H.I.).


\end{document}